\documentclass[aps,prl,reprint,amssymb,amsmath]{revtex4-1}
\usepackage{graphicx}
\usepackage{amsthm}

\def\be{\begin{equation}}
\def\ee{\end{equation}}
\def\ber{\begin{eqnarray}}
\def\eer{\end{eqnarray}}
\def\bern{\begin{eqnarray*}}
\def\eern{\end{eqnarray*}}

\def\rv{\mathbf{r}}

\def\kv{\mathbf{k}}
\def\pv{\mathbf{p}}

\def\0v{\mathbf{0}}
\def\1v{\mathbf{1}}
\def\2v{\mathbf{2}}
\def\3v{\mathbf{3}}

\def\pa{\partial}

\begin{document}

\title{Many-body quantum dynamics by the reduced density matrix based on the time-dependent density functional theory}
\author{Vladimir~U.~Nazarov}
\affiliation{Research Center for Applied Sciences, Academia Sinica, Taipei 11529, Taiwan}
\email{nazarov@gate.sinica.edu.tw}

\begin{abstract}
We  evaluate the density matrix of an arbitrary quantum mechanical system
in terms of  the quantities pertinent to the solution of the time-dependent density functional theory (TDDFT) problem. 
Our theory utilizes the adiabatic connection perturbation method of G\"{o}rling and Levy, 
from which the expansion of the many-body density matrix in powers of the coupling constant $\lambda$ naturally arises.
We then find the reduced density matrix $\rho_\lambda(\rv,\rv',t)$, 
which, by construction, has the $\lambda$-independent diagonal elements
$\rho_\lambda(\rv,\rv,t)=n(\rv,t)$, $n(\rv,t)$ being the particle density. 
The off-diagonal elements of $\rho_\lambda(\rv,\rv',t)$  contribute importantly to the processes,
which cannot be treated  via the density, directly or by the use of the known TDDFT functionals.
Of those, we consider the momentum-resolved photoemission, doing this to the first order in $\lambda$,
i.e., on the level of  the exact exchange theory.
In  illustrative calculations of photoemission from the quasi-2D electron gas 
and isolated atoms, we find quantitatively strong 
and conceptually far-reaching differences
with the independent-particle Fermi's golden rule formula.
\end{abstract}

\maketitle

\footnotetext{See Supplemental Material at $\dots$ for the derivation of (I) Eqs.~(\ref{mat0})-(\ref{mat}), 
(II) Eqs.~(\ref{ph0})-(\ref{def0}),  (III) Eqs.~(\ref{ph1})-(\ref{def3}), and (IV) 
for the reduction of the theory in the case  of the Q2DEG with one filled
subband.}

Time-dependent (TD) density functional theory (TDDFT)  \cite{Zangwill-80,Runge-84,Gross-85} 
is a widely used powerful method to study the time-evolution and 
the excitation processes in quantum mechanical systems.  
Its success is  due to the crucial simplification arising from the substitution of the prohibitively complicated many-body problem  
with the reference single-particle one, keeping (apart from approximations possibly invoked)
the exact TD electron density of the original many-body system.
The description of a number of physical processes (e.g., optical absorption \cite{Kim-02,Botti-04}, 
slowing of ions in matter \cite{Echenique-81,Nazarov-05}, 
impurity resistivity of metals \cite{Nazarov-14-2},
{\it etc.}) can be rigorously reduced to finding the TD electron density, making TDDFT the method of choice for studying those classes of phenomena.

There exist, at the same time, fundamental processes and the corresponding experimental methods, the theory of which cannot,
on the very general physical grounds, be formulated explicitly in terms of the particle density. 
For a clear example, the momentum-resolved
photoemission requires the knowledge of the probability in the {\em momentum space}, which, 
as long as we remain within the framework of the consistent quantum mechanics, cannot be found directly from the probability
in the {\em coordinate space}, the latter giving the particle density. The  necessary information is, in this case, contained in the reduced 
density matrix (DM) $\rho$ \cite{Landau-81}. 
The real space $\rho(\rv,\rv',t)$ and the momentum space $\rho(\pv,\pv',t)$ representations of $\rho$
are related by the double Fourier transform, while  the diagonal elements in the corresponding representations (probabilities) 
cannot be related directly 
\footnote{According to the general principles of TDDFT, all physical quantities, including the momentum distribution, are determined by the particle density. The corresponding functionals are not, however, known, which necessitates such studies as ours.}.

To find the reduced DM is a  complicated problem, generally speaking, taking us back to the many-body theory. 
In this Letter we come up with the observation that the solution of this task can be greatly facilitated if the TDDFT problem for the same system has been already solved. 
We use the power of  the adiabatic connection perturbation method \cite{Gorling-94,Gorling-97} and show that,  
changing the electron-electron ($e$-$e$) interaction constant $\lambda$ continuously 
from zero (for the reference system) to one (for the physical system), 
while keeping the particle density $n_\lambda(\rv)=n(\rv)$ unchanged, 
we determine not only the Kohn-Sham (KS)   \cite{Kohn-65} potential $v_s(\rv,t;\lambda)$, but also the many-body
DM $\hat{\rho}_\lambda$. 
The latter can be readily reduced to the one-DM $\rho_\lambda(\rv,\rv',t)$ expressed through the
KS TDDFT quantities. 
We emphasize, and this is the motivation of this work, that $\rho_{\lambda=1}(\rv,\rv',t)$ is, while
the KS DM  is not, the true reduced DM of the physical system (c.f., Ref.~\cite{Casida-95}).

Practically, the above program can so far be implemented to the first order in $\lambda$ only,
which results in the construction of the TD exact-exchange (TDEXX)-based theory of the DM.
We apply this theory to the problem of the momentum-resolved photoemission, 
finding quantitative and qualitative differences with the Fermi's golden rule.
We use atomic units ($e^2=m_e=\hbar=1$).

{\it Real-time formalism for  DM to the first order in the interaction.}--%
We write the adiabatic connection Hamiltonian for an $N$-particle system \cite{Gorling-94,Gorling-97}
\begin{equation}
\hat{H}(t;\lambda) \! = \! \sum\limits_{i=1}^N \! \left[ -\frac{1}{2} \Delta_i \! + \! v_{ext}(\rv_i,t) \! + \! \tilde{v}(\rv_i,t;\lambda) \right] 
+ \sum\limits_{i<j}^N \frac{\lambda}{|\rv_i \! -\! \rv_j|},
\label{Hac}
\end{equation}
where $\lambda\in[0,1]$, $\tilde{v}(\rv,t;0)=v_s(\rv,t) -v_{ext}(\rv,t)$,
$v_{ext}$ and $v_s$ being the external and KS potentials, respectively,
and we keep the particle density $\lambda$-independent \cite{Gorling-94,Gorling-97}.
The corresponding $N$-body DM  obeys the Liouville's equation
\begin{equation}
i \frac{\pa \hat{\rho}(t;\lambda)}{\pa t}= [\hat{H}(t;\lambda),\hat{\rho}(t;\lambda)].
\label{l}
\end{equation}
Expanding to the first order in $\lambda$ (but making, so far, 
no assumption regarding the strength of the external TD field), we write
\begin{equation}
\left[
\begin{array}{l}
\hat{H}(t;\lambda)\\
\hat{\rho}(t;\lambda)\\
\tilde{v}(t;\lambda)
\end{array}
\right] =
\left[
\begin{array}{l}
\hat{H}_0(t)\\
\hat{\rho}_0(t)\\
\tilde{v}_0(t)
\end{array}
\right]+ \lambda
\left[
\begin{array}{l}
\hat{H}_1(t)\\
\hat{\rho}_1(t)\\
\tilde{v}_1(t)
\end{array}
\right],
\label{v01}
\end{equation}
where
\begin{align}
&\hat{H}_0(t) = \sum\limits_{i=1}^N \left[ -\frac{1}{2} \Delta_i +v_{ext}(\rv_i,t) + \tilde{v}_0(\rv_i,t) \right],\\
&\hat{H}_1(t) = \sum\limits_{i=1}^N \tilde{v}_1(\rv_i,t) 
+ \sum\limits_{i<j}^N \frac{1}{|\rv_i-\rv_j|},\label{H1}
\end{align}
and the corresponding density matrices evolve as
\begin{align}
&i \frac{\pa \hat{\rho}_0(t)}{\pa t}= [\hat{H}_0(t),\hat{\rho}_0(t)], \label{0}\\
&i \frac{\pa \hat{\rho}_1(t)}{\pa t}= [\hat{H}_0(t),\hat{\rho}_1(t)]+[\hat{H}_1(t),\hat{\rho}_0(t)]\label{l1}.
\end{align}
Let for $t\le 0$ the system be in its ground-state
with the KS wave-function $|0\rangle$, where $|\alpha\rangle$ is the orthonormal complete set of the
Slater-determinant eigenfunctions of  $\hat{H}_0(0)$. 
Let at $t=0$ the TD potential be switched on.
Then, since $\hat{H}_0(t)$ is self-conjugate,  $|\alpha(t)\rangle$, which satisfy
\begin{equation}
i \frac{\pa |\alpha(t)\rangle}{\pa t}= \hat{H}_0(t) |\alpha(t)\rangle, \, |\alpha(0)\rangle=|\alpha\rangle,
\label{ts}
\end{equation}
constitute also an orthonormal complete set  at each $t$.
From Eqs.~(\ref{0}) and (\ref{l1}) we obtain \cite[Ref.][Sec. I]{Note1}
\begin{align}
&\langle \alpha(t)| \hat{\rho}_0(t)|  \beta(t) \rangle = \delta_{\alpha 0} \delta_{\beta 0}, 
\label{mat0} \\
&\langle \alpha(t)| \hat{\rho}_1(t)|  \beta(t) \rangle \! = \! i
(\delta_{\alpha 0}-\delta_{\beta 0}) \! \! \! \int\limits_{-\infty}^t \! \! \langle \alpha(t')| \hat{H}_1(t')| \beta(t')\rangle d t',
\label{mat}
\end{align}
where $\delta_{\alpha \beta}$ is the Kronecker symbol.
Transforming Eqs.~(\ref{mat0}) and (\ref{mat}) to real space and reducing to the one-DM,
we find
\begin{align}
&\rho_0(\rv,\rv',t) \! =  \sum\limits_{i\in occ}  \phi_i(\rv,t) \phi_i^*(\rv',t),
 \label{rho0000} \\
&\rho_1(\rv,\rv',t) \! =  \! \! \! \! \! \! \sum\limits_{\substack{i\in occ\\ j\in unocc}} \! \! \! \! \!
 \langle 0(t) | \hat{\rho}_1(t)| 0_{ij}(t)\rangle \phi_i(\rv,t) \phi_j^*(\rv',t)  +  (\rv \! \leftrightarrow \! \rv')^*,
 \label{rho1000}
\end{align}
where $\phi_i$ are KS orbitals, 
$0_{ij}(t)$ is the propagated ground-state Slater-determinant  $0(t)$ with the $i$-th orbital replaced with the $j$-th one
[$\langle 0(t) | \hat{\rho}_1(t)| 0_{ij}(t)\rangle$ are the only matrix elements that survive the integration].
Equation~(\ref{rho1000}) reduces to
\begin{equation}
\begin{split}
&\rho_1(\rv,\rv',t) \! = \! -i \! \! \! \! \sum\limits_{\substack{i\in occ\\ j\in unocc}} \! \! \! \int_{-\infty}^t \! \! \! \! d t'
\left[   \int \! \! v_x(\rv_1,t') \phi_i^*(\rv_1,t')  \phi_j(\rv_1,t') d\rv_1 \right. \\ &\left. 
+    \int \frac{ \phi_i^*(\rv_1,t')  \rho_0(\rv_1,\rv_2,t') \phi_j(\rv_2,t')}{|\rv_1-\rv_2|} d\rv_1 d\rv_2 \right] \phi_i(\rv,t)  \phi_j^*(\rv',t) \\
& + (\rv \leftrightarrow \rv')^*,
\end{split}
\label{rho1}
\end{equation}
where $v_x=v_s-v_{ext}-v_H$,
and $v_H$  are the exchange and the Hartree potentials, respectively.

Setting $\rv'=\rv$ in Eq.~(\ref{rho1}) and equating  to zero 
(the density must be $\lambda$-independent), 
we retrieve the TD version of the optimized effective potential equation \cite{Sharp-53,Talman-76} 
for  $v_x(\rv,t)$.
On the other hand, 
if above  we allowed for nonlocal effective potentials,
then Eq.~(\ref{rho1}) would reproduce the long-known result \cite{Moller-34}
that the Hartree-Fock (HF) potential  nullifies $\rho_1$.
Consequently, the (TD)HF reduced DM is the first-order 
approximation to the physical one. As discussed above, this is not the case within TDDFT.

It is verifiable by the direct substitution that $\rho_0$ of Eq.~(\ref{rho0000}) and $\rho_1$ of Eq.~(\ref{rho1}) satisfy 
the  Liouville-type equations
\begin{align}
&i \frac{\pa \rho_0(\rv,\rv',t)}{\pa t}  = [\hat{h}_s(t),\rho_0(t)] ,
\label{Li0}\\
\begin{split}
&i \frac{\pa \rho_1(\rv,\rv',t)}{\pa t}  = [\hat{h}_s(t),\rho_1(t)] 
  -[ v_x(t), \rho_0(t)]  + \\
&    \int \!  \rho_0(\rv,\rv_1,t)  \rho_0(\rv_1,\rv',t) \left[ \frac{1}{|\rv_1-\rv'|} -\frac{1}{|\rv_1-\rv|}\right]d\rv_1 ,  
\end{split}
\label{Li1}
\end{align}
where $\hat{h}_s(t)$ is the KS  Hamiltonian. 
Equation (\ref{rho1}) or, alternatively, (\ref{Li1}) determine the time-evolution of the reduced DM
to the first order in the  interaction, and they are expected to be useful in the nonlinear dynamics. 
We, however, turn now to the linear response regime and focus on the photoemission spectroscopy (PES) application.

{\it Linear-response theory.}--%
From now on  we assume the TD external potential 
\begin{equation}
v_{ext}^{(1)}(\rv,t) = \frac{1}{2} \left[ v_{ext}^{(1)}(\rv,\omega) e^{-i \omega t} + c.c. \right].
\label{extp}
\end{equation}
to be weak.
We expand
$\rho(t)=\rho^{(0)}+\rho^{(1)}(t)+\rho^{(2)}(t) + \dots$,
where the superscripts stand for the orders in the  strength of the TD perturbation, 
while the subscripts remain reserved for the orders in the $e$-$e$ interaction.
To the zeroth order in the latter, we  obtain for the probability per unit time for an electron to be emitted into the state $\phi_f(\rv)$
\cite[Ref.][Sec. II]{Note1} \footnote{The contribution from $\rho^{(1)}$ term is zero identically.}
\begin{equation}
\lim\limits_{t\to \infty} \! \! \frac{\langle \phi_f| \rho_0^{(2)}(t)|\phi_f\rangle}{t} \! = \!
 \sum\limits_{i\in occ}  
A_{f i}(\omega)  \delta(\omega-\epsilon_f+\epsilon_i),
\label{ph0}
\end{equation}
where
\begin{equation}
A_{f i}(\omega)=
\frac{\pi }{2}
|\langle \phi_f|v_s^{(1)}(\omega)|\phi_i\rangle|^2,
\label{def0}
\end{equation}
which reproduces the conventional Fermi's golden rule.
To the {\em first} order in the interaction,  Eq.~(\ref{Li1}) 
leads to \cite[Ref.][Sec. III]{Note1}
\begin{widetext}
\begin{equation}
\lim\limits_{t\to \infty} \frac{ \langle \phi_f| \rho_1^{(2)}(t)|\phi_f\rangle}{t}   =
 \sum\limits_{i\in occ} \Delta A_{f i}(\omega)  \delta(\omega-\epsilon_f+\epsilon_i) +
 \Delta B_{f i}(\omega)  \delta'(\omega-\epsilon_f+\epsilon_i) ,
\label{ph1}
\end{equation}
\begin{equation}
\begin{split}
& \Delta A_{f i}(\omega)  =  -\pi   \, {\rm Re}   \left \{ \langle \phi_f|v_s^{(1)}(\omega)|\phi_i\rangle^* \left[
   \langle \phi_f|v_x^{(1)}(\omega)|\phi_i\rangle 
+ \sum\limits_{k\ne i} C_{k i}  \frac{\langle \phi_f|v_s^{(1)}(\omega)|\phi_k\rangle }{\epsilon_i-\epsilon_k}     
   \right. \right. \\
& \left. \left.
+   \sum\limits_{k\ne f}
 C_{f k} \frac{ \langle \phi_k|v_s^{(1)}(\omega)|\phi_i\rangle }{\epsilon_f-\epsilon_k}  
  + \sum\limits_{kl} (f_k-f_l) \,
\frac{ \langle \phi_k|v_s^{(1)}(\omega)|\phi_l\rangle}{\epsilon_k-\epsilon_l-\omega-i\eta}  
 \int  \frac{\phi_i(\rv) \phi_f^*(\rv')   \phi_l^*(\rv) \phi_k(\rv') }{|\rv-\rv'|} d\rv   d\rv'   \right] \right\},
\end{split}
\label{def1}
\end{equation}
\end{widetext}
\begin{equation}
\Delta B_{f i} (\omega) =  
- \frac{\pi}{2} |\langle \phi_f|v_s^{(1)}(\omega)|\phi_i\rangle|^2 C_{i i},
\label{def2}
\end{equation}
where
\begin{equation}
C_{k m} \! = \!  \langle\phi_k|v_x^{(0)} |\phi_m\rangle
  +  \!   \int  \! \rho_0^{(0)}(\rv,\rv')   \frac{\phi_k^*(\rv) \phi_m(\rv')}{|\rv-\rv'|}d\rv d\rv' ,
 \label{def3}
\end{equation}
and $f_k$ are the orbitals' occupancies.
Equations (\ref{ph1})-(\ref{def3}) 
generalize the Fermi's golden rule,
including interaction to the first order.

The two terms in Eq.~(\ref{ph1}) have distinct physical meaning: 
The one with the delta-function
accounts for the change in the {\em amplitude} of the emission  due to the $e$-$e$ interaction.
The one with the delta-function derivative accounts for the {\em excitation energies shifts}, 
due to the same reason. To  demonstrate this,
we combine Eqs.~(\ref{ph0}) and (\ref{ph1}) as
\begin{widetext}
\begin{equation}
\begin{split}
\lim\limits_{t\to \infty} \frac{ \langle \phi_f| \rho^{(2)}(t)|\phi_f\rangle}{t}  & = 
 \sum\limits_{i\in occ}  \left[ A_{f i}(\omega) \! + \! \Delta A_{f i}(\omega)\right]  \delta(\omega  -\epsilon_f + \epsilon_i)  +
 \Delta B_{f i}(\omega)  \delta'(\omega-\epsilon_f+\epsilon_i)  
\label{ph} 
 \\ 
 &=
 \sum\limits_{i\in occ}   \left[ A_{f i}(\omega)  +  \Delta A_{f i}(\omega)\right]  
 \delta\left[\omega  - \epsilon_f  + \epsilon_i +\Delta \omega_i\right] , 
 \end{split}
 \end{equation}
 \end{widetext}
 where
\begin{equation}
 \Delta \omega_i =\frac{\Delta B_{f i}(\omega)}{A_{f i}(\omega)}= - C_{i i}.
\label{dw}
\end{equation}
We note that the energy-shift (\ref{dw}) is  a ground-state property of the KS system.
We now turn to illustrative calculations.

\begin{figure}[h!]
\hspace{-0.5 cm}
\includegraphics[width= \columnwidth, trim=0 30 45 20, clip=true]{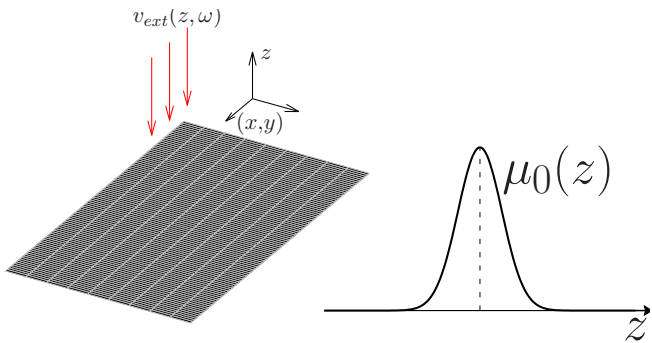}
\caption{\label{sys} Left: Schematics of the Q2DEG with one filled subband. 
Right: Schematics of the wave-function of the only filled subband.}
\end{figure}

{\it Photoemission from quasi-2D electron gas with one filled subband.}--%
For quasi-2D electron gas (Q2DEG) with one filled subband and normally applied electric field 
(schematized in Fig.~\ref{sys}) the analytical solution to the TDEXX problem exists \cite{Nazarov-17},
which makes it ideally suited for the illustration of our theory by  a simple calculation.
Then 
\begin{equation}
\begin{split}
  v_x(z,t)   = -\frac{1}{n_s} \int    \frac{F_2(k_F|z -  z'|)}{|z  - z'|}  n(z',t) d z' ,
\end{split}
\label{main152}
\end{equation} 
where
$F_2(u)=1+[L_1(2 u)-I_1(2 u)]/u$,
$L_1$ and $I_1$ are the 1st-order modified Struve and Bessel functions, 
$n_s=\int_{-\infty}^\infty n(z,t) d z$ is the time-independent 2D density,
and $k_F$ is the corresponding 2D Fermi radius.
From equations (\ref{def0}),  (\ref{def1})-(\ref{def2}), we find $A_{f 0}(\omega)$, $\Delta A_{f 0}(\omega)$,
and $\Delta \omega$  \cite[Ref.][Sec. IV]{Note1}. In particular,
\begin{equation}
\Delta \omega(k_\|) =    
-     \int |\mu_0(z)|^2  G_{k_\|}(z)d z,
\label{shift2D}
\end{equation}
where
$\kv_\|$ is the conserving in-plane momentum, 
\begin{align}
&G_{k_\|}(z)=v_x^{(0)}(z)+ k_F \int |\mu_0(z')|^2 S_{k_\|}(k_F|z-z'|)  d z',\\
&S_{k_\|}(u)= \int\limits_0^\infty \frac{J_1(x) J_0(\frac{k_\|}{k_F} x)}{\sqrt{x^2+u^2}} d x , \label{SSS} 
\end{align}
and $J_n(x)$ are Bessel functions 
\cite[see Ref.~][Sec.~IV for the plot of $S_{k_\|}(u)$]{Note1}. 

\begin{figure}[h!]
\includegraphics[width= \columnwidth, trim=8 0 0 0, clip=true]{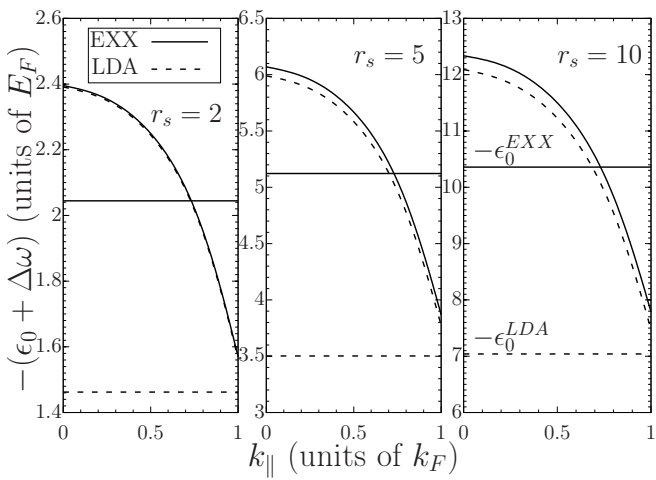}
\caption{\label{dB}
IP obtained with the use of Eq.~(\ref{shift2D}) for Q2DEG with one filled subband,
versus the in-plane momentum, shown
for three values of the density parameter $r_s$.
EXX and LDA-based quantities are plotted with  solid and dashed lines, respectively.
The minus KS eigenvalues $-\epsilon_0$ are shown with  horizontal straight lines. }
\end{figure}

In Fig.~\ref{dB} we plot the ionization potential (IP) of an electron with the momentum $k_\|$. 
The IP with the interactions included (the solid curve for the EXX calculation)
depends on  $k_\|$.
This dependence signifies a fundamental difference between the KS and the many-body dynamics:
Our system is uniform in the $xy$-plane and, therefore, 
$xy$ and $z$ coordinates separate in the KS equations, 
resulting in the motion of a KS electron in the $z$-dimension being
unaffected by the value of its in-plane momentum.
In particular, the IP in the KS dynamics is $k_\|$-independent (shown with horizontal  lines).
Secondly, depending on $k_\|$, the energy shift can be either positive or negative.
Therefore,  for larger $k_\|$, we can emit an electron
with the photon energy $\omega$ less than the KS work function $-\epsilon_0$.
We stress that these results are not in contradiction to the  theorem stating that
the minus highest occupied KS orbital energy is  IP \cite{Perdew-82} (IP-theorem),
since the latter has been proven for finite number of particles
(and then $k_\|$ is not defined), while our case is of infinite number of electrons
\footnote{Interestingly, the IP-theorem holds, in this case, `on average', i.e.,  IP($k_{\|}$), 
averaged over $\kv_{\|}$, equals the minus KS eigenvalue, since $ \int_{k_{\|} \le k_F} \Delta \omega(k_{\|}) d \kv_{\|}=0$,
as can be verified by Eqs.~(\ref{shift2D})-(\ref{SSS}) and (\ref{main152}).}.

Expansion of  DM
in $\lambda$, leading to Eq.~(\ref{rho1}), 
may not necessarily be based on  TDEXX. While the  latter ensures that 
$\rho(\rv,\rv,t)$ is the physical density to the 1st order, 
we could have used other  TDDFT
schemes as well. Then the resulting series  could, likewise,
be expected to converge to the physical DM. 
In Fig.~\ref{dB} we, therefore, compare EXX results   to those of the local density approximation (LDA)
(dashed lines). 
An eloquent conclusion is that, while the KS eigenvalues, being auxiliary quantities,
are completely different in the respective approximations (horizontal lines), the IPs
we obtain, being approximations to physical quantities,
are found close to each other in EXX and LDA. Obviously,
the latter is of great practical consequence, since it shows that inexpensive 
local functionals can be successfully used in the framework of this theory.
\begin{figure}[h!]
\includegraphics[width= \columnwidth, trim=0 0 0 0, clip=true]{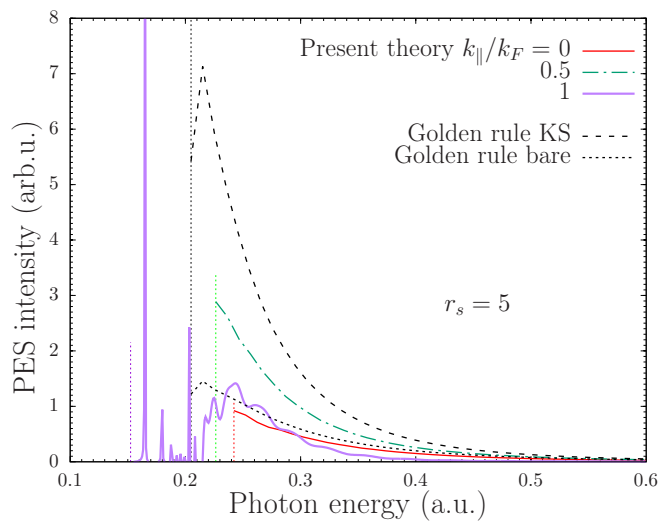}
\caption{\label{MF} Spectra of photoemission from the Q2DEG with one subband filled. 
Results of our theory 
[sum of Eqs.~(\ref{def0}) and (\ref{def1})] are shown with the thin solid red, dashed-dotted green,
and thick solid purple lines, for $k_\|/k_F=0$, $0.5$, and $1$, respectively. Results for the Fermi's golden rule using the KS potential $v^{(1)}_s$ 
[Eq.~(\ref{def0})],
and with the bare external potential $v^{(1)}_{ext}$, are shown with dashed and dotted lines, respectively.
The threshold of photoemission $\omega_{th}=-\epsilon_0-\Delta\omega(k_\|)$ is indicated in each case by a vertical dotted straight line. }
\end{figure}

In Fig.~\ref{MF}, 
we plot  the interacting electrons' emission intensity
and compare it with its Fermi's golden rule counterpart. 
It must be noted 
that  the golden rule 
is overwhelmingly often used in the literature with the KS field in the matrix element replaced with the bare external one (dipole approximation),
while the screening has been included only rather recently \cite{Krasovskii-10}.
It is, therefore, instructive to compare our results to the both variants of the conventional formula. 
Without interaction, the threshold of the photoemission lies at $-\epsilon_0$,
shown in Fig.~\ref{MF} with a long vertical dotted line, and it is the same for all values of $k_\|$.
As discussed above, this is not the case with the interaction  included, 
and the corresponding thresholds for three values of $k_\|$ are shown by short vertical dotted lines.
The spectra at different $k_\|$ are very different from each other, signifying the important
quantitative role of the interaction effect. The case of $k_\|=k_F$ deserves special attention:
Here $\Delta \omega >0$, which makes emission possible at $\omega<-\epsilon_0$.
In this energy range, the spectrum is strongly affected by the  transitions between the ground and discreet excited states,
resulting in resonances at the corresponding energies. Since within TDEXX these transitions are undamped \cite{Nazarov-17},
the amplitudes of the corresponding peaks are not in the same scale with the rest of the spectra.

{\it Isolated atoms.}-- Our second example concerns photoemission from atoms.
In Table \ref{table1} we list the KS EXX eigenvalues,  the energy shifts, and the total
IP according to the present theory. The following important observations can be made.
Firstly, for the highest energy levels, the shifts $\Delta \omega$ disappear, which is in agreement
with the IP-theorem.
\begin{table}[h!]
\caption{KS EXX orbital eigenvalues $\epsilon_i$,
the energy shifts $\Delta \omega_i$ of Eq.~(\ref{dw}), and
the corresponding interaction-corrected IP $-(\epsilon_i + \Delta \omega_i)$ 
for several spherically symmetric spin neutral atoms, compared to the experimental \cite{Shirley-77} and the HF 
\cite{Saito-09} values.}
\centering
\begin{tabular}{r c c c c c} 
 \hline
atom &  $-\epsilon_i$ & $-\Delta \omega_i$ &  $-(\epsilon_i\! + \! \Delta \omega_i$) & $-\epsilon_i^{exp}$  & $-\epsilon_i^{HF}$   \\
 \hline
 He(1s) & 0.9179 & -9.6$\times$10$^{-14}$ & 0.9179 &0.9036& 0.9179  \\ 
 Be(1s) & 4.1147 & 0.6169 & 4.7316 &4.384 &4.7327  \\
 (2s) &  0.3091 & -2.7$\times$10$^{-6}$& 0.3091 & 0.3425 &  0.3093 \\
 Ne(1s) & 30.767& 1.9951 &32.762 &31.985 & 32.772\\
  (2s) &  1.7054 &  0.2187& 1.9241& 1.781 & 1.9304\\
  (2p) &0.8478 & -5.4$\times$10$^{-5}$& 0.8477& 0.7960 & 0.8504 \\
 Mg(1s) & 46.267 &  2.7567 &49.024 & 48.174& 49.032\\
     (2s) & 3.0927 & 0.6697 & 3.7624 & 3.454 & 3.7677\\
     (2p) & 1.8696 & 0.4114 &  2.2811 &  2.0212 & 2.2822  \\
     (3s) & 0.2526 & 3.2$\times$10$^{-5}$ & 0.2526 & 0.2811 & 0.2531  \\
 \hline
\end{tabular}
\label{table1}
\end{table}
Secondly, for inner levels, $\Delta \omega$ are large and they change the KS eigenvalues in the right direction
to the experimental IP. These shifts are, however, too big, 
making the theoretical IP to overestimate the experimental ones,
while the KS values underestimate them.
Obviously, further terms in the series in $\lambda$ are necessary to improve the agreement with experiment.
Thirdly,  our  $\epsilon_i+\Delta \omega_i$
are found very close to the HF eigenvalues.
This has a fundamental reason: 
As follows from the discussion after Eq.~(\ref{rho1}), the latter give physical IP to the first order 
in the  interaction,
which also $\epsilon_i+\Delta \omega_i$ do, but not $\epsilon_i$.

\begin{table}[h!]
\caption{KS LDA and EXX orbital eigenvalues and the corresponding interaction-corrected 
IP of the atoms in Table \ref{table1}.}
\centering
\begin{tabular}{r c c c c c} 
 \hline
atom &  $-\epsilon_i^{LDA}$ &  $-\epsilon_i^{EXX}$  &  $-(\epsilon_i^{LDA} \! + \! \Delta \omega_i^{LDA}$) &$-(\epsilon_i^{EXX} \! + \! \Delta \omega_i^{EXX}$)    \\
 \hline
 He(1s) & 0.5170 & 0.9179 & 0.9354 & 0.9179 \\ 
 Be(1s) & 3.7956 & 4.1147 & 4.7547 &4.7316  \\
 (2s) &  0.1736 &  0.3091 & 0.3123  & 0.3091  \\
 Ne(1s) & 30.229 & 30.767 &32.849 &32.762\\
 (2s) &  1.2656  &  1.7054 & 1.9741& 1.9241 \\
 (2p) &0.4428 &0.8478 & 0.8958&  0.8477  \\
 Mg(1s) & 45.890   & 46.267 &49.090 & 49.024\\
 (2s) & 2.8454  & 3.0927 & 3.7874 & 3.7624\\
 (2p) & 1.6615 & 1.8696 &  2.3102 & 2.2811 \\
 (3s) & 0.1423   & 0.2526 & 0.2542 &  0.2526  \\
 \hline
\end{tabular}
\label{table2}
\end{table}
As seen from Table \ref{table2}, similarly to the case of Q2DEG, the use of LDA instead of EXX does not change
the IP significantly: 
while the orbital eigenvalues differ largely in the corresponding approximations,
adding $\Delta \omega$ brings them close together.

In conclusions,
assuming a solution to the TDDFT problem for a quantum mechanical system  known, 
we have evaluated the reduced density matrix 
$\rho(\rv,\rv',t)$ to the first order in the $e$-$e$ interaction, at the fixed particle density, as stipulated by TDDFT.
The knowledge of $\rho(\rv,\rv',t)$ extends the theory to phenomena, which are now beyond the reach of the pure TDDFT
with the existing observable functionals.
As a particular application, we have derived an extension to the Fermi's golden rule  for the momentum-resolved 
stationary photoelectron  spectroscopy, which accounts for the interparticle  interaction.

Our calculations for the quasi-2D electron gas with one filled subband and for isolated atoms 
manifest an important role of the $e$-$e$  interactions in the TDDFT of  PES. 
In particular, our theory captures a remarkable effect of the correlation between the in-plane and the normal motion 
in a laterally uniform system, which is a feature due to the many-body interactions.

Going beyond the bare exchange
remains the main challenge in the future development of the theory.
Although, on the formal level, our method contains all the correlations at  $\lambda^n, n\ge 2$,
at present only the inclusion of the $\lambda^2$ term looks feasible. 
Since this method involves the TDDFT calculation followed by  the construction of  the reduced DM, 
it comes very encouraging  that, as both our examples show, 
the inaccuracies of the former are compensated by the latter.
This opens the way to use the inexpensive local TDDFT functionals without compromising the accuracy
of the final results, which greatly contributes to the practicability of this method. 

Among other  extensions of the theory, we note that
the nonlinear dynamics using our Eq.~(\ref{Li1}) provides a natural
pathway to the quantum-mechanically 
consistent inclusion of interactions in the theory of photoemission in the time-domain \cite{Pohl-00,DeGiovannini-12,Dinh-13,Dauth-16,Wopperer-17,DeGiovannini-17},
presently this theory relying on the {\em ansatz} of the identification of the KS particles with physical electrons \cite{Dauth-16}.
Finally, we anticipate it conceptually feasible  to extend the theory to evaluate the two-electron density matrix, with an immediate application to the double photoelectron spectroscopy.

\begin{acknowledgments}
Author acknowledges support from the Ministry of Science and Technology, Taiwan, Grants  106--2923--M-001--002--MY3 and 107--2112--M--001--033.
\end{acknowledgments}

\onecolumngrid
\newpage
\thispagestyle{empty}

\renewcommand{\theequation}{{S.\arabic{equation}}}
\renewcommand{\thefigure}{{S.\arabic{figure}}}

\section*{SUPPLEMENTAL MATERIAL}

\begin{center}
to the paper by Vladimir U. Nazarov \\
\ \\
{\bf Many-body quantum dynamics by the TDDFT-based theory of the reduced density matrix}
\end{center}

\setcounter{page}{1}
\setcounter{equation}{0}
\setcounter{figure}{0}
\setcounter{section}{1}

\

\section{(I)  Derivation of  Eqs.~(\ref{mat0}) and (\ref{mat}).}

(a) Taking a matrix element of Eq.~(\ref{0}), we have
\begin{equation}
i \langle \alpha(t)| \frac{\pa \hat{\rho}_0(t)}{\pa t}| \beta(t)\rangle= \langle \alpha(t)| [\hat{H}_0(t),\hat{\rho}_0(t)] | \beta(t)\rangle,
\end{equation}
which, with account of Eq.~(\ref{ts}) leads to
\begin{equation}
i \langle \alpha(t)| \frac{\pa \hat{\rho}_0(t)}{\pa t}| \beta(t)\rangle= 
\langle i \frac{\pa  \alpha(t)}{\pa t} | \hat{\rho}_0(t) | \beta(t)\rangle-
\langle \alpha(t)| \hat{\rho}_0(t)  | i \frac{\pa \beta(t)}{\pa t}\rangle,
\end{equation}
and, therefore, to
\begin{equation}
 \frac{\pa}{\pa t} \langle \alpha(t)| \hat{\rho}_0(t)|  \beta(t) \rangle = 0.
\end{equation}
Hence
\begin{equation}
 \langle \alpha(t)| \hat{\rho}_0(t)|  \beta(t) \rangle = \langle \alpha| \hat{\rho}_0(0)|  \beta \rangle,
\end{equation}
and Eq.~(\ref{mat0}) is proven with account of the fact that at $t=0$ our system is in its ground KS state.

\

(b) Similarly, taking a matrix element of Eq.~(\ref{l1}), we have
\begin{equation}
i \langle \alpha(t)| \frac{\pa \hat{\rho}_1(t)}{\pa t}| \beta(t)\rangle= \langle \alpha(t)| [\hat{H}_0(t),\hat{\rho}_1(t)] | \beta(t)\rangle+ \langle \alpha(t)| [\hat{H}_1(t),\hat{\rho}_0(t)]| \beta(t)\rangle,
\end{equation}
which, with account of Eq.~(\ref{ts}) and of the equation
\begin{equation}
\hat{\rho}_0(t)| \alpha(t)\rangle=\delta_{\alpha 0} |\alpha(t)\rangle
\end{equation}
leads to
\begin{equation}
i \langle \alpha(t)| \frac{\pa \hat{\rho}_1(t)}{\pa t}| \beta(t)\rangle= 
\langle i \frac{\pa  \alpha(t)}{\pa t} | \hat{\rho}_1(t) | \beta(t)\rangle-
\langle \alpha(t)| \hat{\rho}_1(t)  | i \frac{\pa \beta(t)}{\pa t}\rangle
+(\delta_{\beta 0}-\delta_{\alpha 0}) \langle \alpha(t)| \hat{H}_1(t)| \beta(t)\rangle,
\end{equation}
and, therefore, to
\begin{equation}
i \frac{\pa}{\pa t} \langle \alpha(t)| \hat{\rho}_1(t)|  \beta(t) \rangle = 
(\delta_{\beta 0}-\delta_{\alpha 0}) \langle \alpha(t)| \hat{H}_1(t)| \beta(t)\rangle.
\label{lll}
\end{equation}
Equation (\ref{mat}) is obtained by the time integration of Eq.~(\ref{lll}).

\section{(II). Derivation of  Eqs.~(\ref{ph0})-(\ref{def0}).}

We apply the time-dependent perturbation 
\begin{equation}
v_{ext}^{(1)}(t) = \frac{1}{2} \left[ v_{ext}^{(1)}(\omega) e^{-i (\omega+i\eta) t}+
v_{ext}^{(1)}(-\omega) e^{i (\omega-i\eta) t} \right],
\end{equation}
where $\eta$ is a positive infinitesimal, ensuring the perturbation to  be zero at $t\to -\infty$.
Within the linear response, the same holds for the KS potential
\begin{equation}
v_s^{(1)}(t) = \frac{1}{2} \left[ v_s^{(1)}(\omega) e^{-i (\omega+i\eta) t}+
v_s^{(1)}(-\omega) e^{i (\omega-i\eta) t} \right].
\label{Svs}
\end{equation}
Expanding Eq.~(\ref{Li0}) to the second order in the perturbation,
we have
\begin{align}
&i\frac{\pa \rho_0^{(1)}(t)}{\pa t}=[\hat{h}_s^{(0)},\rho_0^{(1)}(t)]+[v_s^{(1)}(t),\rho_0^{(0)}], 
\label{Li01}\\
&i\frac{\pa \rho_0^{(2)}(t)}{\pa t}=[\hat{h}_s^{(0)},\rho_0^{(2)}(t)]+[v_s^{(1)}(t),\rho_0^{(1)}(t)]+[v_s^{(2)}(t),\rho_0^{(0)}].
\label{Li02}
\end{align}
By Eq.~(\ref{Li02}) we have
\begin{equation}
i\frac{\pa \langle \phi_f|\rho_0^{(2)}(t)|\phi_f\rangle}{\pa t}=
\langle\phi_f|[v_s^{(1)}(t),\rho_0^{(1)}(t)]|\phi_f\rangle = 2 i \, {\rm Im} \langle\phi_f|v_s^{(1)}(t) \rho_0^{(1)}(t)|\phi_f\rangle, 
\label{22}
\end{equation}
where $\phi_f$ is the orbital of the emitted electron, and the contributions from the first and the third terms in the right-hand
side of Eq.~(\ref{Li02}) disappear, $\phi_f$ being empty in the ground-state. Equation (\ref{Li01}) gives us
\begin{equation}
\langle \phi_f|\rho_0^{(1)}(t)|\phi_m\rangle= \frac{f_f-f_m}{2}
\left[ \langle \phi_f|v_s^{(1)}(\omega)|\phi_m\rangle \frac{e^{-i( \omega+i\eta) t}}{\epsilon_f-\epsilon_m-\omega-i\eta}  + 
\langle \phi_f|v_s^{(1)}(-\omega)|\phi_m\rangle \frac{e^{i(\omega -i\eta) t}}{\epsilon_f-\epsilon_m+\omega-i\eta} 
\right] .
\label{rho01m}
\end{equation}
Combining Eqs.~(\ref{22}), (\ref{Svs}), and (\ref{rho01m}), we can write
\begin{equation}
\begin{split}
&\frac{\pa \langle \phi_f|\rho_0^{(2)}(t)|\phi_f\rangle}{\pa t}
= 2  \, {\rm Im} \sum\limits_i \langle\phi_f|v_s^{(1)}(t)|\phi_i\rangle \langle \phi_i|\rho_0^{(1)}(t)|\phi_f\rangle = \\
&\frac{1}{2} \, {\rm Im} \! \sum\limits_i (f_i \! - \! f_f) \! 
 \left[ \langle \phi_f|v_s^{(1)}(\omega)|\phi_i \rangle e^{-i (\omega+i\eta) t} \! + \!
\langle \phi_f|v_s^{(1)}(-\omega)|\phi_i \rangle e^{i (\omega-i\eta) t} \right]  \!
 \left[ \! \frac{\langle \phi_i|v_s^{(1)}(\omega)|\phi_f\rangle e^{-i( \omega+i\eta) t}}{\epsilon_i-\epsilon_f-\omega-i\eta}  \! + \!
\frac{\langle \phi_i|v_s^{(1)}(-\omega)|\phi_f\rangle  e^{i(\omega -i\eta) t}}{\epsilon_i-\epsilon_f+\omega-i\eta} 
\! \right] \\
&=\! \frac{e^{2 \eta t}}{2} \, {\rm Im} \! \! \sum\limits_{i\in occ} \!
\frac{|\langle \phi_f|v_s^{(1)}(\omega)|\phi_i \rangle|^2  }{\epsilon_i-\epsilon_f+\omega-i\eta} 
\! + \!
\frac{|\langle \phi_i|v_s^{(1)}(\omega)|\phi_f\rangle |^2}{\epsilon_i-\epsilon_f-\omega-i\eta} \! = \!
\frac{\pi}{2} \! \sum\limits_{i\in occ} \!
|\langle \phi_f|v_s^{(1)}(\omega)|\phi_i \rangle|^2 \delta(\epsilon_i \! - \! \epsilon_f \! +\! \omega)
\! + \!
|\langle \phi_i|v_s^{(1)}(\omega)|\phi_f\rangle |^2 \delta(\epsilon_i \! - \! \epsilon_f \! - \! \omega).
\end{split}
\label{222}
\end{equation}
In the third line of Eq.~(\ref{222}) we have kept the non-oscillating terms only,
and after the last equality sign we have taken the $\eta\to$ limit.
Assuming $\omega>0$ and noting that $\epsilon_f>\epsilon_i$, we conclude the proof of Eqs.~(\ref{ph0})-(\ref{def0}).

\section{(III). Derivation of  Eqs.~(\ref{ph1})-(\ref{def3}).}

We write down  the second-order  term in the expansion of  Eq.~(\ref{Li1}) in powers of the perturbation
\begin{equation}
\begin{split}
&\frac{\pa \langle \phi_f| \rho_1^{(2)}(t)|\phi_f\rangle }{\pa t}  \! =  2 \, {\rm Im} \left\{ -\langle \phi_f| \rho_1^{(1)}(t) v_s^{(1)}(t) |\phi_f\rangle+
 \langle \phi_f| v_s^{(2)}(t) \rho_1^{(0)} |\phi_f\rangle
   + \langle \phi_f| \rho_0^{(1)}(t) v_x^{(1)}(t)|\phi_f\rangle  \right. \\
 & \left.
 +     \sum\limits_{m} \langle \phi_f|\rho_0^{(2)}(t)|\phi_m\rangle  \left[ \langle \phi_m|v_x^{(0)}|\phi_f\rangle +\int  \frac{\phi_m^*(\rv)  \phi_f(\rv')  \rho_0^{(0)}(\rv,\rv')} {|\rv-\rv'|} d\rv  d\rv'\right]  \right. \\
 & \left. +    \sum\limits_{mkl} \langle \phi_f|\rho_0^{(1)}(t)|\phi_m\rangle   \langle \phi_k|\rho_0^{(1)}(t)|\phi_l\rangle 
 \int  \frac{\phi_m^*(\rv) \phi_f(\rv')   \phi_k(\rv) \phi_l^*(\rv') }{|\rv-\rv'|} d\rv   d\rv' \right\}.  
\end{split}
\label{IIIstart}
\end{equation}
In the following, we evaluate term by term in Eq.~(\ref{IIIstart}). In resulting expressions, we retain the non-oscillating terms only,
keeping in view that the oscillating ones do not give a contribution to the final result. Accordingly, we use 
the $\sim$ (tilde) sign to denote the right-hand sides with the oscillating parts dropped. A caution should, however,
be exercised to omit an oscillating term only when it would not be further multiplied by another oscillating one,
yielding a non-oscillating result. For example, in Eq.~(\ref{T1}), oscillating expressions are omitted, while
it would be incorrect to omit such parts in $\rho_1^{(1)}(t)$ before evaluating its product with $v_s^{(1)}(t)$.
All the quantities below are obtained by expanding Eqs.~(\ref{Li0}) or (\ref{Li1}) to the corresponding orders in the time-dependent
perturbation. We arrive at
\begin{equation}
\begin{split}
&\langle \phi_f| \rho_1^{(1)}(t) v_s^{(1)}(t) |\phi_f\rangle \sim
 \frac{1}{4}  \sum\limits_{m\in occ}
 \left[  \frac{\langle \phi_f|v_x^{(1)}(\omega)|\phi_m\rangle \langle \phi_m|v_s^{(1)}(-\omega)|\phi_f\rangle }{\epsilon_f-\epsilon_m-\omega-i\eta} +
\frac{\langle \phi_f|v_x^{(1)}(-\omega)|\phi_m\rangle  \langle \phi_m|v_s^{(1)}(\omega)|\phi_f\rangle}{\epsilon_f-\epsilon_m+\omega-i\eta} \right] + \\
& \! \! \!  \! \!  \! \!  \! \! \! \! \frac{1}{4} \! \sum\limits_{m k} \! \frac{f_k \! - \! f_m}{\epsilon_k \! - \! \epsilon_m}
\! \! \left[ \! \frac{\langle \phi_f|v_s^{(1)}(\omega)|\phi_k\rangle \langle \phi_m|v_s^{(1)}(-\omega)|\phi_f\rangle }{\epsilon_f-\epsilon_m-\omega-i\eta} \! + \!
\frac{\langle \phi_f|v_s^{(1)}(-\omega)|\phi_k\rangle  \langle \phi_m|v_s^{(1)}(\omega)|\phi_f\rangle }{\epsilon_f-\epsilon_m+\omega-i\eta} \! \right]  \!  \! \left[ \! \langle\phi_k|v_x^{(0)} |\phi_m\rangle
\!  +     \! \! \int  \! \! \! \rho_0^{(0)}(\rv,\rv')   \frac{\phi_k^*(\rv) \phi_m(\rv')}{|\rv'-\rv|}d\rv' d\rv  \right] \! \! -  \\
& \! \! \!  \! \!  \! \!  \! \! \! \!  \frac{1}{4} \! \sum\limits_{m k}
 \frac{f_f \! - \! f_k}{\epsilon_f \! - \! \epsilon_k} \! \left[ \! \langle\phi_f|v_x^{(0)} |\phi_k\rangle
\! +   \!  \! \int  \! \! \! \rho_0^{(0)}(\rv,\rv')   \frac{\phi_f^*(\rv) \phi_k(\rv')}{|\rv'-\rv|}d\rv' d\rv  \right] \! \!
\left[ \! \frac{\langle \phi_k|v_s^{(1)}(\omega)|\phi_m\rangle \langle \phi_m|v_s^{(1)}(-\omega)|\phi_f\rangle }{\epsilon_f-\epsilon_m-\omega-i\eta} \! + \!
\frac{\langle \phi_k|v_s^{(1)}(-\omega)|\phi_m\rangle  \langle \phi_m|v_s^{(1)}(\omega)|\phi_f\rangle }{\epsilon_f-\epsilon_m+\omega-i\eta} \! \right] \! \! -
\\
 & \! \! \!  \! \!  \! \!  \! \! \! \!  \sum\limits_{\substack{m\in occ \\ k l}} 
\frac{f_l-f_k}{4}
\left[  \frac{\langle \phi_k|v_s^{(1)}(\omega)|\phi_l\rangle \langle \phi_m|v_s^{(1)}(-\omega)|\phi_f\rangle }{(\epsilon_f \! - \! \epsilon_m \! - \! \omega \! - \! i\eta)(\epsilon_k \! - \! \epsilon_l \! - \! \omega \! - \! i\eta)}  + 
 \frac{\langle \phi_k|v_s^{(1)}(-\omega)|\phi_l\rangle  \langle \phi_m|v_s^{(1)}(\omega)|\phi_f\rangle }{( \epsilon_f \! - \! \epsilon_m \! + \! \omega \! - \! i\eta) (\epsilon_k \! - \! \epsilon_l \! + \! \omega \! - \! i\eta)} 
\right] 
 \int   \frac{\phi_k(\rv') \phi_l^*(\rv) \phi_f^*(\rv') \phi_m(\rv) }{|\rv'-\rv|}  d\rv'   d\rv  \, + \\
 &   \! \! \!  \! \!  \! \!  \! \! \! \! \sum\limits_{m k} 
\! \frac{f_k\! - \! f_f}{4} \! \! 
\left[  \! \frac{\langle \phi_f|v_s^{(1)}(\omega)|\phi_k\rangle \langle \phi_m|v_s^{(1)}(-\omega)|\phi_f\rangle }{(\epsilon_f \! - \!\epsilon_m \! - \! \omega \! - \! i\eta) (\epsilon_f \! - \! \epsilon_k \! - \! \omega \! - \! i\eta)}  \! + \! 
 \frac{\langle \phi_f|v_s^{(1)}(-\omega)|\phi_k\rangle  \langle \phi_m|v_s^{(1)}(\omega)|\phi_f\rangle}{(\epsilon_f \! - \! \epsilon_m \! + \! \omega \! - \! i\eta)(\epsilon_f \! - \! \epsilon_k \! + \! \omega \! - \! i\eta)} \!
\right]   \! \!
 \left[ \! \langle\phi_k|v_x^{(0)} |\phi_m\rangle \! + \! \! \int   \! \! \rho_0^{(0)}(\rv,\rv') \frac{\phi_k^*(\rv) \phi_m(\rv')}{|\rv'-\rv|}   d\rv' d\rv \right] \! \! - \\
 &    \! \! \!  \! \!  \! \!  \! \! \! \! \sum\limits_{m k}  
\! \frac{f_m \! - \! f_k}{4} \! \! 
\left[ \frac{ \langle \phi_k|v_s^{(1)}(\omega)|\phi_m\rangle \langle \phi_m|v_s^{(1)}(-\omega)|\phi_f\rangle }{(\epsilon_f \! - \! \epsilon_m \! - \!  \omega \! -i\eta)(\epsilon_k \! - \! \epsilon_m \! - \! \omega \! - \! i\eta)}  \! + \! 
\frac{\langle \phi_k|v_s^{(1)}(-\omega)|\phi_m\rangle   \langle \phi_m|v_s^{(1)}(\omega)|\phi_f\rangle }{(\epsilon_f \! - \! \epsilon_m \! + \! \omega \! - \! i\eta) (\epsilon_k \! - \! \epsilon_m \! + \! \omega \! - \! i\eta)} 
\right] \! \! 
 \left[ \! \langle\phi_f|v_x^{(0)} |\phi_k\rangle \! + \! \! \int   \rho_0^{(0)}(\rv,\rv')  \frac{\phi_f^*(\rv) \phi_k(\rv') }{|\rv'-\rv|}  d\rv'   d\rv \right] \! ,
\end{split}
\label{T1}
\end{equation}

\begin{equation}
\langle \phi_f| v_s^{(2)}(t) \rho_1^{(0)} |\phi_f\rangle\sim \sum\limits_{m\in occ}
\frac{\langle \phi_f| v_s^{(2)}(0) |\phi_m\rangle}{\epsilon_f-\epsilon_m} \left[ \langle\phi_m|v_x^{(0)} |\phi_f\rangle
 +    \int  \rho_0^{(0)}(\rv,\rv')   \frac{\phi_m^*(\rv) \phi_f(\rv')}{|\rv-\rv'|}d\rv d\rv'  \right],
 \label{T2}
\end{equation}
where we have used the following two equations
\begin{equation}
v_s^{(2)}(t) = v_s^{(2)}(2 \omega) e^{2 (-i \omega+ \eta) t}+
v_s^{(2)}(0) e^{2\eta t}+
v_s^{(2)}(-2 \omega)e^{2 (i \omega+\eta) t} ,
\end{equation}
\begin{equation}
\begin{split}
& \langle\phi_n|\rho_1^{(0)}|\phi_m\rangle =
  \frac{f_m-f_n}{\epsilon_n-\epsilon_m} \left[ \langle\phi_n|v_x^{(0)} |\phi_m\rangle
\! +   \!  \int  \rho_0^{(0)}(\rv,\rv_1)   \frac{\phi_n^*(\rv) \phi_m(\rv_1)}{|\rv_1-\rv|}d\rv_1 d\rv  \right].
\end{split}
\end{equation}

\begin{equation}
\langle \phi_f| \rho_0^{(1)}(t) v_x^{(1)}(t)|\phi_f\rangle \sim
- \frac{1}{4} \sum\limits_{m\in occ}
\left[  \frac{\langle \phi_f|v_s^{(1)}(\omega)|\phi_m\rangle \langle \phi_m|v_x^{(1)}(-\omega)|\phi_f\rangle}{\epsilon_f-\epsilon_m-\omega-i\eta}  + 
 \frac{\langle \phi_f|v_s^{(1)}(-\omega)|\phi_m\rangle \langle \phi_m|v_x^{(1)}(\omega)|\phi_f\rangle}{\epsilon_f-\epsilon_m+\omega-i\eta} 
\right],
\end{equation}

\begin{equation}
\begin{split}
\langle \phi_f|\rho_0^{(2)}(t)|\phi_m\rangle \sim
\frac{1}{4} \sum\limits_k  \left[ (f_m-f_k)  \frac{\langle \phi_f|v_s^{(1)}(\omega)|\phi_k\rangle 
\langle \phi_k|v_s^{(1)}(-\omega)|\phi_m\rangle}{(\epsilon_f-\epsilon_m-2 i\eta)(\epsilon_k-\epsilon_m+\omega-i\eta)} 
 + 
(f_m-f_k)
 \frac{\langle \phi_f|v_s^{(1)}(-\omega)|\phi_k\rangle 
 \langle \phi_k|v_s^{(1)}(\omega)|\phi_m\rangle}{(\epsilon_f-\epsilon_m-2 i\eta) (\epsilon_k-\epsilon_m-\omega-i\eta)} \right. \\
\left. -  (f_k-f_f)
\frac{\langle \phi_f|v_s^{(1)}(\omega)|\phi_k\rangle \langle \phi_k|v_s^{(1)}(-\omega)|\phi_m\rangle}{(\epsilon_f-\epsilon_m-2i\eta) (\epsilon_f-\epsilon_k-\omega-i\eta)}  
-  (f_k-f_f)
 \frac{\langle \phi_f|v_s^{(1)}(-\omega)|\phi_k\rangle \langle \phi_k|v_s^{(1)}(\omega)|\phi_m\rangle}{(\epsilon_f-\epsilon_m -2 i\eta)(\epsilon_f-\epsilon_k+\omega-i\eta)} \right]
+(f_f-f_m) 
\frac{\langle \phi_f|v_s^{(2)}(0)|\phi_m\rangle}{\epsilon_f-\epsilon_m-i\eta},
\end{split}
\label{T4}
\end{equation}

\begin{equation}
\sum\limits_{mkl} \langle \phi_f|\rho_0^{(1)}(t)|\phi_m\rangle   \langle \phi_k|\rho_0^{(1)}(t)|\phi_l\rangle 
 \!  \sim \! \frac{1}{4}\sum\limits_{\substack{m\in occ \\ k l}} \!  (f_l-f_k) \! 
\left[ \frac{\langle \phi_f|v_s^{(1)}(\omega)|\phi_m\rangle  \langle \phi_k|v_s^{(1)}(-\omega)|\phi_l\rangle}{(\epsilon_f \! - \! \epsilon_m \! - \! \omega \! - \! i\eta)(\epsilon_k \! - \! \epsilon_l \! + \! \omega \! - \! i\eta)}  \! + \!
 \frac{\langle \phi_f|v_s^{(1)}(-\omega)|\phi_m\rangle \langle \phi_k|v_s^{(1)}(\omega)|\phi_l\rangle}{(\epsilon_f \! - \! \epsilon_m \! +\! \omega \! - \! i\eta)(\epsilon_k \! - \! \epsilon_l \! - \! \omega \! - \! i\eta)} 
\right].
\label{T5}
\end{equation}

After Eqs.~(\ref{T1})-(\ref{T5}) are substituted into Eq.~(\ref{IIIstart}), a number of simplifications occur,
and, remembering that $\eta$ is a positive infinitesimal,
we arrive at Eqs.~(\ref{ph1})-(\ref{def3}). The most part of the transformations being straightforward, we mention only the 
two keypoints:

(I) The two instances of the quadratic KS potential $v_s^{(2)}$, which are present in Eqs.~(\ref{T2}) and (\ref{T4}), cancel each other 
(note, that $\epsilon_f>\epsilon_m$ in the last term of Eq.~(\ref{T4}), so that $ i \eta$ can be dropped from its denominator).
The latter is a very fortunate development, since the evaluation of the quadratic response would have presented an insurmountable
difficulty for nontrivial systems; 

(II) The demonstration of the emergence of the delta-function derivative in Eq.~(\ref{ph1}) is  nontrivial, and we, therefore, 
give some additional details. The origin lies in the second and the fifth terms in Eq.~(\ref{T1}). We write
\begin{equation}
\begin{split}
& \! \! \!  \! \!  \! \!  \! \! \! \! Q=\frac{1}{4} \! \sum\limits_{m k} \! \frac{f_k \! - \! f_m}{\epsilon_k \! - \! \epsilon_m \pm i \eta}
\! \! \left[ \! \frac{\langle \phi_f|v_s^{(1)}(\omega)|\phi_k\rangle \langle \phi_m|v_s^{(1)}(-\omega)|\phi_f\rangle }{\epsilon_f-\epsilon_m-\omega-i\eta} \! + \!
\frac{\langle \phi_f|v_s^{(1)}(-\omega)|\phi_k\rangle  \langle \phi_m|v_s^{(1)}(\omega)|\phi_f\rangle }{\epsilon_f-\epsilon_m+\omega-i\eta} \! \right]  C_{k m} +  \\
 &   \! \! \!  \! \!  \! \!  \! \! \! \! \frac{1}{4} \sum\limits_{m k} 
 (f_k\! - \! f_f)\! \! 
\left[  \! \frac{\langle \phi_f|v_s^{(1)}(\omega)|\phi_k\rangle \langle \phi_m|v_s^{(1)}(-\omega)|\phi_f\rangle }{(\epsilon_f \! - \!\epsilon_m \! - \! \omega \! - \! i\eta) (\epsilon_f \! - \! \epsilon_k \! - \! \omega \! - \! i\eta)}  \! + \! 
 \frac{\langle \phi_f|v_s^{(1)}(-\omega)|\phi_k\rangle  \langle \phi_m|v_s^{(1)}(\omega)|\phi_f\rangle}{(\epsilon_f \! - \! \epsilon_m \! + \! \omega \! - \! i\eta)(\epsilon_f \! - \! \epsilon_k \! + \! \omega \! - \! i\eta)} \!
\right]   C_{k m},
\end{split}
\end{equation}
where $C_{k m}$ is given by Eq.~(\ref{def3}). Then
\begin{equation}
\begin{split}
& Q=\frac{1}{4} \! \sum\limits_{m k} \! \frac{f_k \! - \! f_f}{\epsilon_k \! - \! \epsilon_m \pm i \eta}
\! \! \left[ \! \frac{\langle \phi_f|v_s^{(1)}(\omega)|\phi_k\rangle \langle \phi_m|v_s^{(1)}(-\omega)|\phi_f\rangle }{\epsilon_f-\epsilon_m-\omega-i\eta} \! + \!
\frac{\langle \phi_f|v_s^{(1)}(-\omega)|\phi_k\rangle  \langle \phi_m|v_s^{(1)}(\omega)|\phi_f\rangle }{\epsilon_f-\epsilon_m+\omega-i\eta} \! \right]  C_{k m} +  \\
& \frac{1}{4} \! \sum\limits_{m k} \! \frac{f_f \! - \! f_m}{\epsilon_k \! - \! \epsilon_m \pm i \eta}
\! \! \left[ \! \frac{\langle \phi_f|v_s^{(1)}(\omega)|\phi_k\rangle \langle \phi_m|v_s^{(1)}(-\omega)|\phi_f\rangle }{\epsilon_f-\epsilon_m-\omega-i\eta} \! + \!
\frac{\langle \phi_f|v_s^{(1)}(-\omega)|\phi_k\rangle  \langle \phi_m|v_s^{(1)}(\omega)|\phi_f\rangle }{\epsilon_f-\epsilon_m+\omega-i\eta} \! \right]  C_{k m} +  \\
 & \frac{1}{4} \sum\limits_{m k} 
 (f_k\! - \! f_f)\! \! 
\left[  \! \frac{\langle \phi_f|v_s^{(1)}(\omega)|\phi_k\rangle \langle \phi_m|v_s^{(1)}(-\omega)|\phi_f\rangle }{(\epsilon_f \! - \!\epsilon_m \! - \! \omega \! - \! i\eta) (\epsilon_f \! - \! \epsilon_k \! - \! \omega \! - \! i\eta)}  \! + \! 
 \frac{\langle \phi_f|v_s^{(1)}(-\omega)|\phi_k\rangle  \langle \phi_m|v_s^{(1)}(\omega)|\phi_f\rangle}{(\epsilon_f \! - \! \epsilon_m \! + \! \omega \! - \! i\eta)(\epsilon_f \! - \! \epsilon_k \! + \! \omega \! - \! i\eta)} \!
\right]   C_{k m}.
\end{split}
\end{equation}
Summing up the first and the third terms, we have
\begin{equation}
\begin{split}
& Q=\frac{1}{4}  \sum\limits_{m k} \! \frac{f_k \! - \! f_f}{\epsilon_k \! - \! \epsilon_m \pm i \eta}
\! \! \left[ \! \frac{\langle \phi_f|v_s^{(1)}(\omega)|\phi_k\rangle \langle \phi_m|v_s^{(1)}(-\omega)|\phi_f\rangle }{\epsilon_f-\epsilon_k-\omega-i\eta} \! + \!
\frac{\langle \phi_f|v_s^{(1)}(-\omega)|\phi_k\rangle  \langle \phi_m|v_s^{(1)}(\omega)|\phi_f\rangle }{\epsilon_f-\epsilon_k+\omega-i\eta} \! \right]  C_{k m} +  \\
&\frac{\pm i\eta}{4}  \sum\limits_{m k} \! \frac{f_k \! - \! f_f}{\epsilon_k \! - \! \epsilon_m \pm i \eta}
\! \! \left[ \! \frac{\langle \phi_f|v_s^{(1)}(\omega)|\phi_k\rangle \langle \phi_m|v_s^{(1)}(-\omega)|\phi_f\rangle }{(\epsilon_f-\epsilon_k-\omega-i\eta) (\epsilon_f-\epsilon_m-\omega-i\eta)} \! + \!
\frac{\langle \phi_f|v_s^{(1)}(-\omega)|\phi_k\rangle  \langle \phi_m|v_s^{(1)}(\omega)|\phi_f\rangle }{(\epsilon_f-\epsilon_k+\omega-i\eta)(\epsilon_f-\epsilon_m+\omega-i\eta)} \! \right]  C_{k m} + \\
& \frac{1}{4} \sum\limits_{m k} \! \frac{f_f \! - \! f_m}{\epsilon_k \! - \! \epsilon_m \pm i \eta}
\! \! \left[ \! \frac{\langle \phi_f|v_s^{(1)}(\omega)|\phi_k\rangle \langle \phi_m|v_s^{(1)}(-\omega)|\phi_f\rangle }{\epsilon_f-\epsilon_m-\omega-i\eta} \! + \!
\frac{\langle \phi_f|v_s^{(1)}(-\omega)|\phi_k\rangle  \langle \phi_m|v_s^{(1)}(\omega)|\phi_f\rangle }{\epsilon_f-\epsilon_m+\omega-i\eta} \! \right]  C_{k m}.
\end{split}
\end{equation}
Interchanging the dummy $k$ and $m$ indices  in the first term, we have 
\begin{equation}
\begin{split}
& Q=\frac{1}{4}  \sum\limits_{m k} \! \frac{f_f \! - \! f_m}{\epsilon_k \! - \! \epsilon_m \mp i \eta}
\! \! \left[ \! \frac{\langle \phi_f|v_s^{(1)}(\omega)|\phi_m\rangle \langle \phi_k|v_s^{(1)}(-\omega)|\phi_f\rangle }{\epsilon_f-\epsilon_m-\omega-i\eta} \! + \!
\frac{\langle \phi_f|v_s^{(1)}(-\omega)|\phi_m\rangle  \langle \phi_k|v_s^{(1)}(\omega)|\phi_f\rangle }{\epsilon_f-\epsilon_m+\omega-i\eta} \! \right]  C_{m k} +  \\
&\frac{\pm i\eta}{4}  \sum\limits_{m k} \! \frac{f_k \! - \! f_f}{\epsilon_k \! - \! \epsilon_m \pm i \eta}
\! \! \left[ \! \frac{\langle \phi_f|v_s^{(1)}(\omega)|\phi_k\rangle \langle \phi_m|v_s^{(1)}(-\omega)|\phi_f\rangle }{(\epsilon_f-\epsilon_k-\omega-i\eta) (\epsilon_f-\epsilon_m-\omega-i\eta)} \! + \!
\frac{\langle \phi_f|v_s^{(1)}(-\omega)|\phi_k\rangle  \langle \phi_m|v_s^{(1)}(\omega)|\phi_f\rangle }{(\epsilon_f-\epsilon_k+\omega-i\eta)(\epsilon_f-\epsilon_m+\omega-i\eta)} \! \right]  C_{k m} + \\
& \frac{1}{4} \sum\limits_{m k} \! \frac{f_f \! - \! f_m}{\epsilon_k \! - \! \epsilon_m \pm i \eta}
\! \! \left[ \! \frac{\langle \phi_f|v_s^{(1)}(\omega)|\phi_k\rangle \langle \phi_m|v_s^{(1)}(-\omega)|\phi_f\rangle }{\epsilon_f-\epsilon_m-\omega-i\eta} \! + \!
\frac{\langle \phi_f|v_s^{(1)}(-\omega)|\phi_k\rangle  \langle \phi_m|v_s^{(1)}(\omega)|\phi_f\rangle }{\epsilon_f-\epsilon_m+\omega-i\eta} \! \right]  C_{k m},
\end{split}
\end{equation}
or
\begin{equation}
\begin{split}
& Q=
\frac{\pm i\eta}{4}  \sum\limits_{m k} \! \frac{f_k \! - \! f_f}{\epsilon_k \! - \! \epsilon_m \pm i \eta}
\! \! \left[ \! \frac{\langle \phi_f|v_s^{(1)}(\omega)|\phi_k\rangle \langle \phi_m|v_s^{(1)}(-\omega)|\phi_f\rangle }{(\epsilon_f-\epsilon_k-\omega-i\eta) (\epsilon_f-\epsilon_m-\omega-i\eta)} \! + \!
\frac{\langle \phi_f|v_s^{(1)}(-\omega)|\phi_k\rangle  \langle \phi_m|v_s^{(1)}(\omega)|\phi_f\rangle }{(\epsilon_f-\epsilon_k+\omega-i\eta)(\epsilon_f-\epsilon_m+\omega-i\eta)} \! \right]  C_{k m} + \\
& \frac{1}{2} \sum\limits_{m k} \! (f_f \! - \! f_m)
  {\rm Re} \left[ \frac{\langle \phi_f|v_s^{(1)}(\omega)|\phi_k\rangle \langle \phi_m|v_s^{(1)}(-\omega)|\phi_f\rangle C_{k m}}{\epsilon_k \! - \! \epsilon_m \pm i \eta } \right] \frac{1}{\epsilon_f-\epsilon_m-\omega-i\eta}   + \\
&  \frac{1}{2} \sum\limits_{m k} \! (f_f \! - \! f_m)
{\rm Re} \left[ \frac{\langle \phi_f|v_s^{(1)}(-\omega)|\phi_k\rangle  \langle \phi_m|v_s^{(1)}(\omega)|\phi_f\rangle C_{k m}}{\epsilon_k \! - \! \epsilon_m \pm i \eta } \right] \frac{1}{\epsilon_f-\epsilon_m+\omega-i\eta}.
\end{split}
\end{equation}
Then
\begin{equation}
\begin{split}
& {\rm Im \, Q}=
\frac{1}{4} {\rm Im}  \sum\limits_{m } (f_m \! - \! f_f)
\! \! \left[ \! \frac{|\langle \phi_f|v_s^{(1)}(\omega)|\phi_m\rangle|^2  }{ (\epsilon_f-\epsilon_m-\omega-i\eta)^2} \! + \!
\frac{| \langle \phi_m|v_s^{(1)}(\omega)|\phi_f\rangle|^2 }{(\epsilon_f-\epsilon_m+\omega-i\eta)^2} \! \right]  C_{m m} + \\
& \frac{1}{2}  {\rm Im} \sum\limits_{m k} \! (f_f \! - \! f_m)
  {\rm Re} \left[ \frac{\langle \phi_f|v_s^{(1)}(\omega)|\phi_k\rangle \langle \phi_m|v_s^{(1)}(-\omega)|\phi_f\rangle C_{k m}}{\epsilon_k \! - \! \epsilon_m \pm i \eta } \right] \frac{1}{\epsilon_f-\epsilon_m-\omega-i\eta}   + \\
&  \frac{1}{2}  {\rm Im}  \sum\limits_{m k} \! (f_f \! - \! f_m)
{\rm Re} \left[ \frac{\langle \phi_f|v_s^{(1)}(-\omega)|\phi_k\rangle  \langle \phi_m|v_s^{(1)}(\omega)|\phi_f\rangle C_{k m}}{\epsilon_k \! - \! \epsilon_m \pm i \eta } \right] \frac{1}{\epsilon_f-\epsilon_m+\omega-i\eta},
\end{split}
\end{equation}
or
\begin{equation}
\begin{split}
& {\rm Im \, Q}=
-\frac{\pi}{4} {\rm Im}  \sum\limits_{m } (f_m \! - \! f_f) |\langle \phi_f|v_s^{(1)}(\omega)|\phi_m\rangle|^2    C_{m m} 
\delta' (\epsilon_f-\epsilon_m-\omega)+ \\
& \frac{\pi}{2}  \sum\limits_{m \ne k} \! (f_f \! - \! f_m)
  {\rm Re} \left[ \frac{\langle \phi_f|v_s^{(1)}(\omega)|\phi_k\rangle \langle \phi_m|v_s^{(1)}(-\omega)|\phi_f\rangle C_{k m}}{\epsilon_k \! - \! \epsilon_m  } \right] \delta(\epsilon_f-\epsilon_m-\omega) ,
\end{split}
\end{equation}
where we have used the relations
\begin{align}
&\lim\limits_{\eta\to 0} {\rm Im} \, \frac{1}{x- i\eta} =\pi \delta(x),\\
&\lim\limits_{\eta\to 0} {\rm Im} \, \frac{1}{(x- i\eta)^2} =-\pi \delta'(x),
\end{align}
and dropped the terms with $\delta(\epsilon_f-\epsilon_m+\omega)$ and $\delta'(\epsilon_f-\epsilon_m+\omega)$, since $\epsilon_f \in unocc$ and, hence, $\epsilon_m\in occ$, and $\omega$ is assumed positive. 

\section{(IV). Reduction of Eqs.~(\ref{def0}),  (\ref{def1})-(\ref{def2}) in the case  of the Q2DEG with one filled
subband.}

As shown below, in the specific case of the Q2DEG with one subband filled, 
equations (\ref{def0}),  (\ref{def1})-(\ref{def2}) reduce to
\begin{equation}
A_{f 0}(\omega) =\frac{\pi}{2} H(k_F-k_\|)  
\left|\langle \mu_f |v_s^{(1)}(\omega)|\mu_0)\rangle\right|^2 
\label{FGRS}
\end{equation}
and
\begin{equation}
\begin{split}
&\Delta A_{f 0}(\omega)=   
-\pi H(k_F-k_\|)  \, {\rm Re}   \left\{ \langle \mu_f|v_s^{(1)}(\omega)|\mu_0\rangle^* \left[  \frac{k_F}{n_s}  \!
 \int  \mu_0(z')  \mu_n^*(z')    n^{(1)}(z,\omega)   S_{k_\|}(|z-z'|) d z d z' 
 \right. \right. \\
& \left. 
 + \langle \mu_f|v_x^{(1)}(\omega)|\mu_0\rangle  
+  \frac{1}{2 n_s}   \int \frac{\mu_n^*(z)}{\mu_0(z)} v_s^{(1)}(z, \omega) \chi_s(z,z') G_{k_\|}(z') d z d z' 
+   \frac{1}{\omega}\langle \mu_0|v_s^{(1)}(\omega)|\mu_0\rangle  \langle\mu_f|G_{k_\|} |\mu_0\rangle  \right.  \\
& \left. \left. + \frac{1}{2 n_s} 
\int
\frac{\mu_n^*(z) }{\mu_0(z)}  n^{(1)}(z,\omega)    G_{k_\|}(z)  d z 
 - \frac{\omega}{(2 n_s)^2} 
\int \frac{ \mu_n^*(z')  }{|\mu_0(z)|^2 \mu_0(z')} \chi_s(z,z')
  n^{(1)}(z,\omega)     G_{k_\|}(z')  d z d z'  \right] \right\},
\end{split}
\label{amp1S}
\end{equation}
\begin{equation}
\Delta \omega(k_\|) =    
-     \int |\mu_0(z)|^2  G_{k_\|}(z)d z,
\label{shift2DS}
\end{equation}
where
$H(k)$ is the Heaviside step function, 
$n^{(1)}(z,\omega)$ is the density fluctuation,
$\mu_m(z)$ are the orbitals of the perpendicular motion, 
$\lambda_m$ are the corresponding eigenenergies, $\kv_\|$ is the conserving parallel wave-vector, 
common for the initial and final KS states,
$\chi_s(z,z')$ is the static KS density response function. 

The derivation of Eqs.~(\ref{FGRS}), (\ref{amp1S}), and (\ref{shift2DS}) is as follows.
KS orbitals are (for brevity, we omit the `parallel' index in $\pv_\|$)
\begin{equation}
\phi_{m \pv}(\rv)= \frac{1}{\sqrt{\Omega}} e^{i \pv\cdot \rv_\|} \mu_m(z),
\end{equation}
where $\Omega$ is the normalization area, and the eigenenergies corresponding to $\mu_m(z)$ will
be denoted by $\lambda_m$. Only the orbitals
\begin{equation}
\phi_{0 \pv}(\rv)= \frac{1}{\sqrt{\Omega}} e^{i \pv \cdot \rv_\|} \mu_0(z), \ \  |\pv | \le k_F,
\end{equation}
are  occupied. We then evaluate in a straightforward manner
\begin{equation}
\begin{split}
C_{\pv k, \pv' m}=
 \delta_{\pv \pv'} \left[  \langle \mu_k|v_x^{(0)}| \mu_m\rangle+
 k_F \int \mu_0(z) \mu_0^*(z') \mu_k^*(z) \mu_m(z') S_p(k_F |z-z'|)  dz dz'  \right],
\end{split}
\end{equation}
where the function $S$ is given by Eq.~(\ref{SSS}). Furthermore, remembering that $i\in occ$ and $f\in unocc$,
we find
\begin{equation}
\begin{split}
&\sum\limits_{l=1}^\infty C_{l 0}  \frac{\langle \mu_f|v_s^{(1)}(\omega)|\mu_l\rangle }{\lambda_0-\lambda_l} =
\sum\limits_{l=1}^\infty \frac{\langle \mu_f|v_s^{(1)}(\omega)|\mu_l\rangle }{\lambda_0-\lambda_l}   \langle \mu_l|G_{k_\|}| \mu_0\rangle=
 \\
& \int \sum\limits_{l=1}^\infty \frac{\mu_f^*(z') v_s^{(1)}(z',\omega) \mu_l(z')\mu_l(z) G_{k_\|}(z)  \mu_0(z)} {\lambda_0-\lambda_l} d z d z'=
\frac{1}{2 n_s} \int \frac{\mu_f^*(z') }{\mu_0(z')}  v_s^{(1)}(z',\omega)  \chi_s(z',z)  G_{k_\|}(z)  d z d z',
\end{split}
\label{N10}
\end{equation}
\begin{equation}
\begin{split}
&  \sum\limits_{kl} (f_k \! - \! f_l) 
\frac{ \langle \mu_k|v_s^{(1)}(\omega)|\mu_l\rangle}{\epsilon_k \! - \! \epsilon_l \! - \! \omega \! - \! i\eta}  
\!  \int \!  \frac{\phi_i(\rv) \phi_f^*(\rv')   \phi_l^*(\rv) \phi_k(\rv') }{|\rv-\rv'|} d\rv   d\rv'   \! =  \!
 k_F \! \sum\limits_{l=1}^\infty \! \left[
\frac{ \langle \mu_0|v_s^{(1)}(\omega)|\mu_l\rangle}{\lambda_0 \! - \! \lambda_l \! -\omega \! -i\eta}  
\!  \int  \! \! S_{k_\|}(k_F |z \! - \! z'|) \mu_0(z) \mu_f^*(z') \mu_l(z) \mu_0(z') d z d z'   \right.
 \\
 & \left.
  - 
\frac{ \langle \mu_l|v_s^{(1)}(\omega)|\mu_0\rangle}{\lambda_l \! -\lambda_0 \! -\omega \! -i\eta}  
 \! \int  \!  \! S_{k_\|}(k_F |z \! - \! z'|) \mu_0(z) \mu_f^*(z') \mu_l(z') \mu_0(z)
d z d z'   \right],
\end{split}
\label{N11}
\end{equation}
\begin{equation}
\begin{split}
\sum\limits_l
 C_{f l} \frac{ \langle \mu_l|v_s^{(1)}(\omega)|\mu_0\rangle }{\omega+i\eta +\lambda_0-\lambda_l} =
\sum\limits_l
\left[ \langle \mu_f|v_x^{(0)}|\mu_l\rangle+k_F \int \mu_0(z) \mu_0^*(z') \mu_f^*(z) \mu_l(z') S_p(k_F |z-z'|)  dz dz' \right]
\frac{ \langle \mu_l|v_s^{(1)}(\omega)|\mu_0\rangle }{\omega+i\eta +\lambda_0-\lambda_l}.
\label{N12}
\end{split}
\end{equation}
From Eqs.~(\ref{N11}) and (\ref{N12}) we have
\begin{equation}
\begin{split}
&  \sum\limits_{kl} (f_k-f_l) \,
\frac{ \langle \mu_k|v_s^{(1)}(\omega)|\mu_l\rangle}{\epsilon_k-\epsilon_l-\omega-i\eta}  
 \int  \frac{\phi_i(\rv) \phi_f^*(\rv')   \phi_l^*(\rv) \phi_k(\rv') }{|\rv-\rv'|} d\rv   d\rv'  
 +\sum\limits_l
 C_{f l} \frac{ \langle \mu_l|v_s^{(1)}(\omega)|\mu_0\rangle }{\omega+i\eta +\lambda_0-\lambda_l}  
 = \\
&  \frac{k_F}{n_s} 
\!  \int  \! S_{k_\|}(k_F|z \! - \! z'|) \chi_s(z,z'',\omega) v_s^{(1)}(z'',\omega) \mu_f^*(z') \mu_0(z')   d z d z' dz'' 
\! + \! \sum_l \frac{ \langle \mu_l|v_s^{(1)}(\omega)|\mu_0\rangle }{\omega\! + \! i\eta \! + \! \lambda_0 \! - \! \lambda_l} 
\langle \mu_f|G_{k_\|}|\mu_l\rangle 
= \\
&  \frac{k_F}{n_s} 
\int  S_{k_\|}(k_F|z  -  z'|) n^{(1)}(z,\omega) \mu_f^*(z') \mu_0(z')   d z d z' 
 +  \sum_l \frac{ \langle \mu_l|v_s^{(1)}(\omega)|\mu_0\rangle }{\omega +  i\eta  +  \lambda_0  -  \lambda_l} 
\langle \mu_f|G_{k_\|}|\mu_l\rangle
= \\
&  \frac{k_F}{n_s} 
\int  S_{k_\|}(k_F|z  -  z'|) n^{(1)}(z,\omega) \mu_f^*(z') \mu_0(z')   d z d z' 
 +  \frac{ \langle \mu_0|v_s^{(1)}(\omega)|\mu_0\rangle }{\omega} 
\langle \mu_f|G_{k_\|}|\mu_0\rangle
+  \sum_{l=1}^\infty \frac{ \langle \mu_l|v_s^{(1)}(\omega)|\mu_0\rangle }{\omega +  i\eta  +  \lambda_0  -  \lambda_l} 
\langle \mu_f|G_{k_\|}|\mu_l\rangle
= \\
&  \frac{k_F}{n_s} \!
\int  \! S_{k_\|}(k_F|z  \! - \! z'|) n^{(1)}(z,\omega) \mu_f^*(z') \mu_0(z')   d z d z' 
 \! +  \! \frac{ \langle \mu_0|v_s^{(1)}(\omega)|\mu_0\rangle }{\omega} 
\langle \mu_f|G_{k_\|}|\mu_0\rangle
\! +  \! \frac{1}{n_s} \! \int \!  \frac{\mu_f^*(z)}{\mu_0(z)} G_{k_\|}(z)  \tilde{\chi}(z,z',\omega) v_s^{(1)}(z',\omega) 
d z d z',
\label{N13}
\end{split}
\end{equation}
where 
\begin{equation}
\chi_s(z,z',\omega) = n_s \mu_0(z) \mu_0(z') \sum\limits_{l=1}^\infty 
\left( \frac{1}{\omega+i\eta+\lambda_0-\lambda_l} + \frac{1}{-\omega-i\eta+\lambda_0-\lambda_l} \right) \mu_l(z) \mu_l(z'),
\label{chis}
\end{equation}
is the density-response function of the Q2DEG with one filled subband \cite{Nazarov-17}, and
in the last line of Eq.~ (\ref{N13}) we have introduced the notation
\begin{equation}
\tilde{\chi}_s(z,z',\omega) = n_s \mu_0(z) \mu_0(z') \sum\limits_{l=1}^\infty \frac{\mu_l(z) \mu_l(z')}{\omega+i\eta+\lambda_0-\lambda_l}.
\end{equation}

The proof of Eq.~(\ref{amp1S}) is concluded by summing up Eqs.~(\ref{N10}) and (\ref{N13}) and noting that
\begin{equation}
\tilde{\chi}_s(z,z',\omega) = \frac{1}{2} \chi_s(z,z',\omega) -\frac{\omega}{4 n_s} \int \frac{\chi_s(z,z'') \chi_s(z'',z',\omega)}{|\mu_0(z'')|^2} d z''.
\end{equation}
Finally, the latter equality is proven by the direct substitution of Eq.~(\ref{chis}) into the second term on its right-hand side and 
the integration,
taking into account  the orthonormality of $\mu_l(z)$.

\begin{figure}[h!]
\includegraphics[width= 0.7 \columnwidth, trim=0 0 10 0, clip=true]{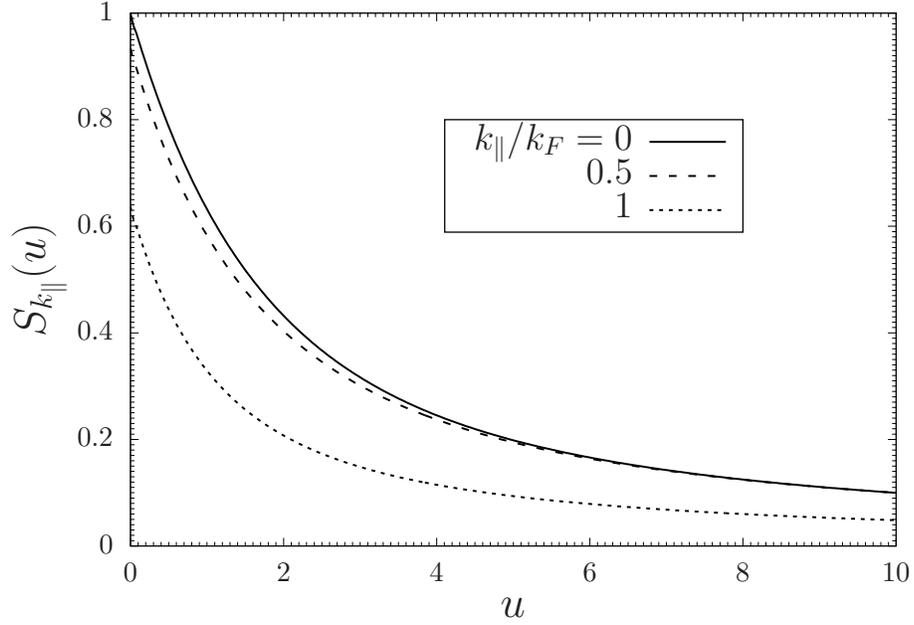}
\caption{\label{FSSS} Function $S_{k_\|}(u)$ of Eq.~(\ref{SSS}) for three values of the in-plane momentum $k_\|$.}
\end{figure}

Function  $S_{k_\|}(u)$ of Eq.~(\ref{SSS}) is plotted in Fig.~\ref{FSSS}.

\end{document}